\documentclass[aip, 
jcp, longbibliography, 
twocolumn, floatfix, superscriptaddress, reprint, 10pt]{revtex4-1}

\usepackage[dvipsnames]{xcolor}
\usepackage[english]{babel}
\usepackage[utf8]{inputenc}
\usepackage[paperwidth=210mm,paperheight=297mm,centering,hmargin=2cm,vmargin=2.5cm]{geometry}
\usepackage[utf8]{inputenc}
\usepackage{verbatim}
\usepackage{comment}
\usepackage{bm}
\usepackage{amsmath}
\usepackage{amsfonts}
\usepackage{natbib}
\usepackage{pdfpages}
\usepackage{graphics}
\usepackage{float}
\usepackage{hyperref}
\usepackage[capitalize]{cleveref}
\usepackage{multirow}
\usepackage{booktabs}
\usepackage{makecell}
\usepackage{outlines}
\usepackage{tabularx}
\usepackage[version=4]{mhchem}
\usepackage{ulem}

\usepackage{multirow}

\newcolumntype{C}{>{\centering\arraybackslash}X}

\usepackage{physics}
\usepackage{xparse}

\newcommand{\mbf}[1]{\ensuremath{\mathbf{#1}}}

\NewDocumentCommand{\rep}{s d<| d|>}{%
\IfBooleanTF{#1}{
   \IfValueTF{#2}{
       \IfValueTF{#3}{\braket{#2}{#3}}{\bra{#2}}
       }{
       \IfValueTF{#3}{\ket{#3}}{}
       }
   }{
   \IfValueTF{#2}{
       \IfValueTF{#3}{\braket*{#2}{#3}}{\bra*{#2}}
       }{
       \IfValueTF{#3}{\ket*{#3}}{}
       }
   }
}

\NewDocumentCommand{\rbra}{sm}{\IfBooleanTF{#1}{\rep*<#2|}{\rep<#2|}}
\NewDocumentCommand{\rket}{sm}{\IfBooleanTF{#1}{\rep*|#2>}{\rep|#2>}}
\NewDocumentCommand{\rbraket}{smm}{\IfBooleanTF{#1}{\rep*<#2||#3>}{\rep<#2||#3>}}

\NewDocumentCommand{\field}{o m e{_} e{^} o e{_} e{^}}{
\IfValueTF{#5}{\overline{
  #2\IfValueT{#3}{_#3}\IfValueT{#4}{^{\otimes #4}} %
  \otimes 
  #5\IfValueT{#6}{_#6}\IfValueT{#7}{^{\otimes #7}} %
  \IfValueT{#1}{;#1}
}}{
  \IfValueTF{#4}{\overline{
     #2\IfValueT{#3}{_#3}\IfValueT{#4}{^{\otimes #4}}
     \IfValueT{#1}{;#1}
  }}
  {#2\IfValueT{#3}{_#3}}
}
}

\NewDocumentCommand{\frho}{o e{_} e{^}}{
\field[#1]{\rho}_{#2}^{#3}
}
\NewDocumentCommand{\fdelta}{o e{_} e{^}}{
\field[#1]{\delta}_{#2}^{#3}
}

\newcommand{\e}{a}  %

\newcommand{\br}{\mbf{r}}
\newcommand{\bx}{\mbf{x}}

\newcommand{\brhat}{\hat{\mbf{r}}}

\NewDocumentCommand{\ex}{e_}{
\IfValueTF{#1}{\e_{#1}\bx_{#1}}{\e\bx}
}  %

\NewDocumentCommand{\lm}{e_}{
\IfValueTF{#1}{l_{#1}m_{#1}}{lm}
}
\NewDocumentCommand{\nlm}{e_}{
\IfValueTF{#1}{n_{#1}\lm_{#1}}{n\lm}
}
\NewDocumentCommand{\enlm}{e_}{
\IfValueTF{#1}{\e_{#1}\nlm_{#1}}{\e\nlm}
}
\NewDocumentCommand{\en}{e_}{
\IfValueTF{#1}{\e_{#1}n_{#1}}{\e n}
}
\NewDocumentCommand{\nlk}{e_}{
\IfValueTF{#1}{n_{#1}l_{#1}k_{#1}}{nlk}
}
\NewDocumentCommand{\enlk}{e_}{
\IfValueTF{#1}{\e_{#1}\nlk_{#1}}{\e\nlk}
}
\NewDocumentCommand{\enl}{e_}{
\IfValueTF{#1}{\en_{#1}l_#1}{\en l}
}
\NewDocumentCommand{\nl}{e_}{
\IfValueTF{#1}{n_{#1}l_#1}{n l}
}

\NewDocumentCommand{\nnl}{s}{
\IfBooleanTF{#1}{n_1 n_2 l}{n_1; n_2; l}
}
\NewDocumentCommand{\ennl}{s}{
\IfBooleanTF{#1}{\en_1 \en_2 l}{\en_1; \en_2; l}
}

\NewDocumentCommand{\gslm}{s}{
\IfBooleanTF{#1}{\sigma\lambda\mu}{\sigma;\lambda\mu}
}

\newcommand{\lmax}{l_\text{max}}

\usepackage{pdfpages}
\usepackage{pgffor}
\makeatletter
\AtBeginDocument{\let\LS@rot\@undefined}
\makeatother
\begin{document}

\setcitestyle{super}

\title{Fast evaluation of spherical harmonics with {\texttt{sphericart}}}

\author{Filippo Bigi}
\author{Guillaume Fraux}
\affiliation{Laboratory of Computational Science and Modelling, Institute of Materials, {\'E}cole Polytechnique F\'ed\'erale de Lausanne, Lausanne 1015, Switzerland}
\affiliation{National Centre for Computational Design and Discovery of Novel Materials (MARVEL), {\'E}cole Polytechnique F{\'e}d{\'e}rale de Lausanne, 1015 Lausanne, Switzerland}

\author{Nicholas J. Browning}
\affiliation{Swiss National Supercomputing Centre (CSCS), 6900, Lugano, Switzerland}

\author{Michele Ceriotti}
\affiliation{Laboratory of Computational Science and Modelling, Institute of Materials, {\'E}cole Polytechnique F\'ed\'erale de Lausanne, Lausanne 1015, Switzerland}
\affiliation{National Centre for Computational Design and Discovery of Novel Materials (MARVEL), {\'E}cole Polytechnique F{\'e}d{\'e}rale de Lausanne, 1015 Lausanne, Switzerland}
\affiliation{michele.ceriotti@epfl.ch}

\onecolumngrid
\begin{abstract}
Spherical harmonics provide a smooth, orthogonal, and symmetry-adapted basis to expand functions on a sphere, and they are used routinely in physical and theoretical chemistry as well as in different fields of science and technology, from geology and atmospheric sciences to signal processing and computer graphics.
More recently, they have become a key component of rotationally equivariant models in geometric machine learning, including applications to atomic-scale modeling of molecules and materials.
We present an elegant and efficient algorithm for the evaluation of the real-valued spherical harmonics. 
Our construction features many of the desirable properties of existing schemes and allows to compute Cartesian derivatives in a numerically stable and computationally efficient manner. 
To facilitate usage, we implement this algorithm in {\texttt sphericart}, a fast C++ library which also provides C bindings, a Python API, and a PyTorch implementation that includes a GPU kernel.
\end{abstract}
\twocolumngrid
\maketitle

\section{Introduction}

The spherical harmonics $Y^m_l$ are basis functions for the irreducible representations of the SO(3) group,\cite{clausmller_1966_spherical} which makes them a key tool in understanding physical phenomena that exhibit rotational symmetries, and to design practical algorithms to model them on a computer. 
Examples include the distribution of charge in atoms\cite{steiner1963charge}, the behavior of gravitational\cite{rexer_2016_layerbased} and magnetic\cite{knaack_2005_spherical, morschhauser_2014_a} fields, and the propagation of light\cite{evans_1998_the} and sound\cite{poletti2005three-dimensional} in the atmosphere. 
Moreover, they provide a complete set of smooth, orthogonal functions defined on the surface of a sphere, and in this sense they are widely used in many fields including computer graphics\cite{max1988spherical, sloan2008stupid}, quantum~\cite{schlegel1995transformation, varganov2008resolutions, gill2009resolutions, maintz2016efficient} and physical~\cite{pere-yang96jcp,choi+99jcp,ding+17jcp} chemistry, and signal processing\cite{zotkin2009regularized, li2011spherical}.
More recently, they have become an essential tool in the context of geometric deep learning\cite{cohe-well16icml,bron+21arxiv}, as a structural descriptor needed in SO(3)-, O(3)-, and E(3)-equivariant machine learning models\cite{thom+18arxiv,ande+19nips,klic+21arxiv,geiger2022e3nn} and more specifically in the construction of symmetry-adapted descriptors of atomic structures in chemical machine learning\cite{bart+13prb,thom+15jcp,will+19jcp,drau19prb,musi+21cr}.  
Derivatives of the spherical harmonics are also very often used in these applications. An example is that of the calculation of forces on a configuration of atoms by an E(3)-invariant machine learning model.\cite{christensen2020role} Likewise, spherical harmonics gradients have been used in computer graphics for mid-range illumination, irradiance volumes, and optimization methods.\cite{sloan2008stupid}

In many of these contexts, the spherical harmonics are used together with an expansion in the radial direction, to compute expressions of the form
\begin{equation}
c_{nlm} = \sum_k R_{nl}(r_k) Y^m_l(\hat{\br}_k),
\label{eq:expansion}
\end{equation}
where $R_{nl}$ indicates a radial basis, and $\br_k=r_k\brhat_k$ a set of points that correspond to either particles, or to the positions of a Cartesian mesh. 
In others, such as message-passing neural networks,\cite{schu+21icml} directional terms that depend on the orientation of interatomic separation vectors are multiplied by continuous filters that are a function of the radial distance. 
For this reason, although the spherical harmonics are most often defined as complex functions in spherical coordinates, in practical implementations it is often preferred to use their real-valued combinations, and to express their value and derivatives directly in terms of the Cartesian coordinates of the points at which they are evaluated. 
With these applications in mind, we derive compact, efficient expressions to compute the real spherical harmonics and their derivatives of arbitrary order as polynomials of the Cartesian coordinates, we discuss the computational implications of this formulation, and we present a simple yet efficient implementation that can be used both as a C and C++ library and as a Python module.

\section{Analytical expressions}

The real-valued spherical harmonics can be defined in spherical coordinates $(\theta,\phi)$ as 
\begin{equation}\label{eq:real-sh-spherical}
    Y_l^m(\theta, \phi) = F_l^{|m|} * P_l^{|m|}(\cos{\theta}) *
    \begin{cases}
      \sin{(|m|\phi)} & \text{if $m < 0$}\\
      1/\sqrt{2} & \text{if $m = 0$}\\
      \cos{(m\phi)} & \text{if $m > 0$}
    \end{cases}      
\end{equation}
where $P_l^m$ is an associated Legendre polynomial and $F_l^m$ is a prefactor which takes the form
\begin{equation}
F_l^m = (-1)^m \sqrt{\frac{2l+1}{2\pi}\frac{(l-m)!}{(l+m)!}}.
\label{eq:Flm}
\end{equation}
Possible strategies for a stable and efficient computation of the $F_l^m$ are discussed in Appendix \ref{app:Flm}.

Traditionally\cite{press07book}, the real spherical harmonics of points in 3D space have been calculated by first converting the Cartesian coordinates $x$, $y$, and $z$ into the spherical coordinates $\theta$ and $\phi$, and then using Eq. \ref{eq:real-sh-spherical} to calculate the spherical harmonics, most often via the standard recurrence relations for $P_l^m(t)$:
\begin{equation}
\begin{split}
    &P_0^0 = 1, \quad    
    P_l^l = - \, (2l-1) \, \sqrt{1-t^2} \, P_{l-1}^{l-1}, \\
    &P_l^{l-1} = (2l-1) \, t \, P_{l-1}^{l-1},\\
    &P_l^m = [(2l-1)\,t\,P_{l-1}^m - (l+m-1)\,P_{l-2}^m]/(l-m).
\end{split}\label{eq:recur-p}
\end{equation}
It should be noted how naive differentiation of these recursions introduces poles on the z axis where $t = \cos{\theta} = \pm 1$, as well as numerical instabilities in the calculation of the derivatives for points close to the z axis. 

In contrast, our algorithm aims at calculating $\tilde{Y}_l^m = r^l Y_l^m$, where $r = \sqrt{x^2+y^2+z^2}$. These are the so-called \emph{solid} harmonics, which consist of simple homogeneous polynomials of the Cartesian coordinates.
This choice avoids the need to normalize $r$, and it leads to simple and stable iterations to compute $\tilde{Y}_l^m$, as well as very compact expressions for their derivatives with respect to the Cartesian coordinates which re-use the same factors needed to evaluate $\tilde{Y}_l^m$. 
In most applications that require spherical harmonics in Cartesian coordinates, the radial direction is dealt with by a separate expansion (cf. Eq.~\eqref{eq:expansion}), and the $r^l$ factor that is included in the scaled $\tilde{Y}_l^m$ can be compensated for at little to no additional cost by a corresponding $r^{-l}$ factor in the radial term.
If, instead, the spherical harmonics are needed in their conventional version, they can be recovered easily from the radially scaled (or solid) kind.

As shown in Appendix A, the scaled Cartesian harmonics $\tilde{Y}_l^m$ can be computed as
\begin{equation}\label{eq:real-sh-cartesian}
    \tilde{Y}_l^m(x, y, z) = F_l^{|m|} Q_l^{|m|}(z, r) \times
    \begin{cases}
      s_{|m|}(x, y) & \text{if $m < 0$}\\
      1/\sqrt{2} & \text{if $m = 0$}\\
      c_m(x, y) & \text{if $m > 0$}
    \end{cases}     
\end{equation}
where we define 
\begin{equation}
\begin{split}
    Q_l^m =& r^l r_{xy}^{-m} \, P_l^m, \\
    s_m =& r_{xy}^m \, \sin{(m\phi)}, \\
    c_m =& r_{xy}^m \, \cos{(m\phi)},
\end{split}\label{eq:define-q}
\end{equation}
with $r_{xy} = \sqrt{x^2+y^2}$. 
Similar (but not equivalent) definitions have been used often in the literature, e.g. in Refs. \citenum{sloa13jcgt,pere-yang96jcp,drau20prb}. The quantities $Q_l^m$, $c_m$, and $s_m$  can be evaluated very efficiently by recursion in Cartesian coordinates. For example, the recursions for $Q_l^m$ follow almost immediately (see Appendix A) from those for $P_l^m$ (Eq.~\eqref{eq:recur-p}) and the definition of $Q_l^m$ (Eq. \eqref{eq:define-q}):
\begin{equation}
\begin{split}
&Q_0^0 = 1, \quad  Q_l^l = - \, (2l-1) \, Q_{l-1}^{l-1}, \\
&Q_l^{l-1} = (2l-1) \, z \, Q_{l-1}^{l-1}=-z Q_l^l, \\
&Q_l^m = [(2l-1)\,z\,Q_{l-1}^m - (l+m-1)\,r^2\,Q_{l-2}^m]/(l-m).
\end{split}\label{eq:iter-q}
\end{equation}
Many other recursive expressions can be derived based on analogous, well-known relations for $P_l^m$, e.g.
\begin{equation}
\begin{split}
Q_l^m = \frac{2(m+1)\,z\,Q_l^{m+1}+r_{xy}^2\,Q_l^{m+2}}{(l+m+1)(l-m)}.
\end{split}\label{eq:iter-q-alt}
\end{equation}
which can be used to iterate \emph{down} from $Q_l^l$, avoiding the pole at $r_{xy}=0$ that is present for the similar recursion for $P_l^m$.

Similarly, recursive relations for $s_m$ and $c_m$ can be derived (see Appendix A) as
\begin{equation}
\begin{split}
&s_0 = 0, \quad  c_0 = 1, \\
&s_m = x s_{m-1} + y c_{m-1},  \quad
c_m = - y s_{m-1} + x c_{m-1}.
\end{split}\label{eq:iter-sc}
\end{equation}
Once the $Q_l^m$, $c_m$, and $s_m$ quantities are known, their Cartesian derivatives also follow in an extremely compact form:

\begin{equation}
    \begin{split}    
&\frac{\partial Q_l^m}{\partial x} \!=\! x Q_{l-1}^{m+1}, \ 
\frac{\partial Q_l^m}{\partial y} \!=\! y Q_{l-1}^{m+1}, \ 
\frac{\partial Q_l^m}{\partial z} \!=\! (l+m) Q_{l-1}^m, \\
&\frac{\partial s_m}{\partial x} = m \,s_{m-1}, \ 
\frac{\partial s_m}{\partial y} = m \, c_{m-1},  \  \\
&\frac{\partial c_m}{\partial x} = m \, c_{m-1},  \ 
\frac{\partial c_m}{\partial y} = - \, m \, s_{m-1}.
    \end{split}\label{eq:iter-deri}
\end{equation}
These relationships (proven in Appendix B) are much simpler than the standard recurrence relations for the derivatives of $P_l^m$ (see e.g. Ref. \citenum{alken2022gsl}), and they do not lead to poles or instabilities when combined to compute the derivatives of $\tilde{Y}^m_l$. The  simplicity of the expressions in Eq. \ref{eq:iter-deri} opens the door to the efficient calculation of Cartesian derivatives of $\tilde{Y}^m_l$ of any order.
In addition, one can find several expressions that directly link \emph{some} of the derivatives of the Cartesian spherical harmonics to other $(l,m)$ values, e.g.
\begin{equation}
\begin{split}
\frac{\partial \tilde{Y}^m_l}{\partial z} =({F_l^{|m|}}/{F_{l-1}^{|m|}})(l+m) \tilde{Y}_m^{l-1},\\
\frac{\partial\tilde{Y}_l^l}{\partial x} = -l(2l-1) ({F_l^{l}}/{F_{l-1}^{l-1}}) \tilde{Y}_{l-1}^{l-1}
\end{split}
\end{equation}
that simplify even further the calculation of the derivatives of the spherical harmonics. 
For $l\le 6$, we use a computer algebra system to automatically find the expressions that provide $\tilde{Y}^m_l$ and their derivatives in terms of lower-$l$ values with the smallest number of multiplications, and we use them to generate hard-coded implementations, as discussed in the following Section.

\section{Computer implementation and benchmarking}

Most computational applications require the evaluation of all the spherical harmonics, and possibly their derivatives, for the values of $l$ up to a maximum degree $l_\mathrm{max}$.  In many cases, the spherical harmonics have to be computed for many points simultaneously, e.g. for the interatomic separation vectors of all neighbors of a selected atom.

Even though we recommend to use the scaled spherical harmonics in applications, accounting for the fact that accompanying radial terms have to compensate for the scaling, we also provide an implementation of ``normalized'' spherical harmonics, by simply evaluating $\tilde{Y}_l^m(x/r,y/r,z/r)$ and applying the chain rule to the derivatives. These additional operations typically result in an overhead of 5-10\%{} for the calculation of the spherical harmonics and their derivatives.

We implement a general routine that takes a list of Cartesian coordinates and evaluates the scaled $\tilde{Y}_l^m$ and, optionally, their derivatives. We use C++ for the implementation, using templates to exploit compile-time knowledge of the maximum angular momentum, the need to compute derivatives, etc. We then define a pure C API covering the typical use cases on top of this C++ API. Since almost all programming languages have a way to call C functions, this enables using our code from most languages used in a scientific context. We also provide a Python package with a high-level interface to the C library.

Furthermore, we use the C++ API to provide an implementation compatible with PyTorch\cite{pasz+19nips}, using a custom backward function to compute gradients using the derivatives evaluated in the forward pass, and making sure the code is compatible with TorchScript, allowing to use models without a Python interpreter. Finally, we implement a GPU-accelerated version of the PyTorch code for NVIDIA GPUs, using the CUDA language.

\subsection{CPU implementation details}

We apply a number of trivial (and a few less obvious) optimizations. For example, we pre-compute the factors $F_l^m$ (Eq.~\eqref{eq:Flm}) to minimize the number of operations in the inner loop of the iterative algorithm discussed above. 
In order to further accelerate the evaluation of low-$l$ Cartesian harmonics, we use a computer algebra system to derive expressions that evaluate the full $\tilde{Y}^m_l$ and their derivatives with hard-coded expressions using the smallest possible number of multiplications. 
As shown in Table~\ref{tab:serial-timings}, and as noted in previous implementations\cite{sloa13jcgt, geiger2022e3nn}, there is considerable scope for optimization by using ad hoc expressions. However, the speedup is less remarkable when including the calculation of the derivatives through Eq.~\eqref{eq:iter-deri}, which re-uses quantities that have already been computed when evaluating $\tilde{Y}_l^m$.

\begin{table}[btp]
    \centering
    \renewcommand{\arraystretch}{1.18}
    \begin{tabularx}{\linewidth}{C CC CC}
        \hline\hline
         & \multicolumn{2}{c}{ general-purpose} & \multicolumn{2}{c}{ hard-coded} \\
         & $\tilde{Y}_l^m$ & $\tilde{Y}_l^m, \grad\tilde{Y}_l^m$ & $\tilde{Y}_l^m$ & $\tilde{Y}_l^m, \grad\tilde{Y}_l^m$\\
         \hline
            $l_\text{max}=1$ & 3.49 & 11.0 & 1.33 & 7.85 \\
            $l_\text{max}=2$ & 7.16 & 21.6 & 4.21 & 16.4 \\
            $l_\text{max}=3$ & 13.3 & 36.1 & 7.26 & 28.4 \\
            $l_\text{max}=4$ & 20.3 & 55.3 & 11.8 & 46.3 \\
            $l_\text{max}=5$ & 29.8 & 81.9 & 16.3 & 68.3 \\
            $l_\text{max}=6$ & 42.8 & 121  & 22.2 & 95.2 \\
         \hline\hline
    \end{tabularx}
    \caption{Serial execution time (in ns/point) for computing Cartesian spherical harmonics and their derivatives up to the indicated value of $l_\text{max}$ in double precision on an Intel Xeon Gold 6226R CPU, averaged over 1000 calls for 10~000 points, comparing our general purpose recursive algorithm to an hard-coded implementation.}
    \label{tab:serial-timings}
\end{table}

\begin{figure}[bthp]
    \centering
    \includegraphics[width=1.0\linewidth]{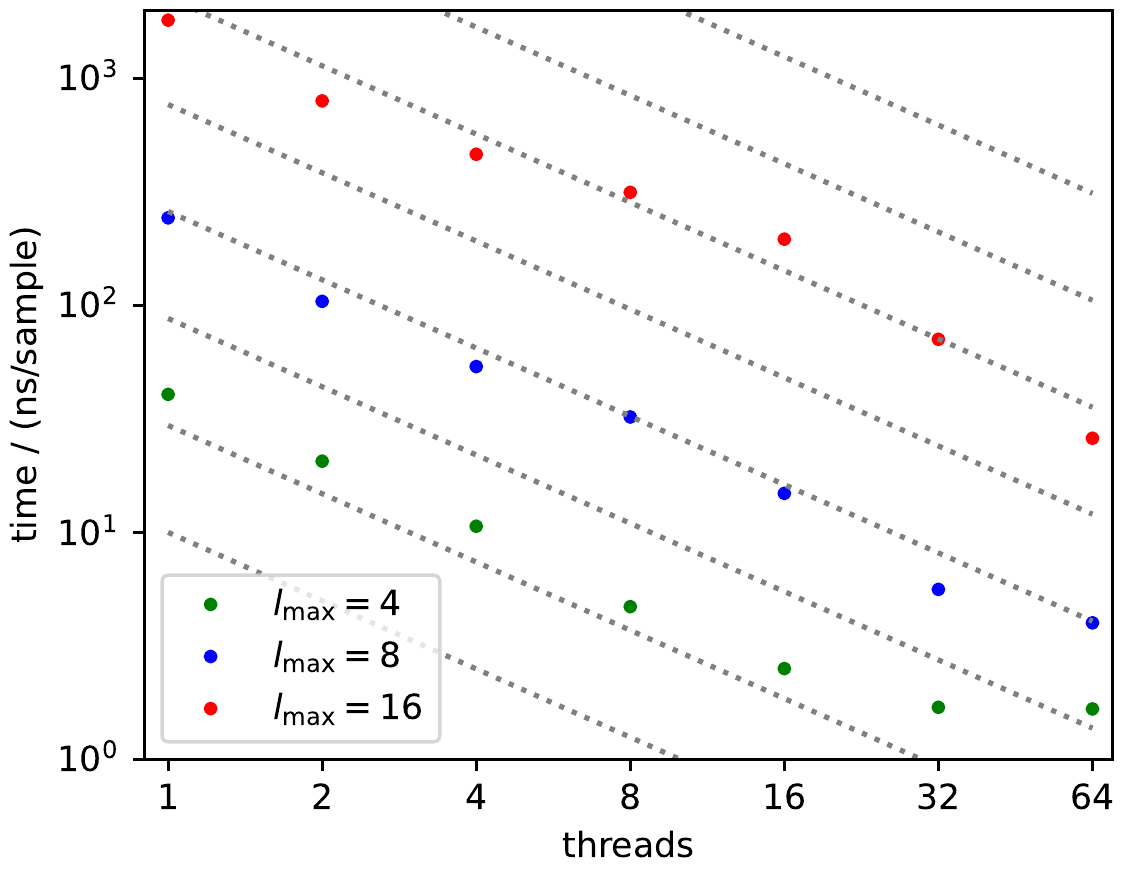}
    \caption{Scaling of wall-clock timing to evaluate the Cartesian spherical harmonics and their derivatives for 10~000 points using different $l_\text{max}$ and numbers of OpenMP threads. Results refer to up to 64 cores of up to two Intel Xeon Platinum 8360Y (2.4 GHz) CPUs.}
    \label{fig:scaling}
\end{figure}

As a compromise between the convenience of a function that works for arbitrary $l_\text{max}$ and the efficiency of optimized expressions, we provide an interface for the hard-coded implementation up to $l_\text{max}=6$, and use by default a hybrid implementation that applies the hard-coded version for small $l$, and then switches to the general expression, using the alternative recursion~\eqref{eq:iter-q-alt} to avoid computing low-$l$ values of the modified Legendre polynomials $Q_l^m$. 
All our functions can be applied to many 3D points at once, and they are trivially parallelized over the sample index using OpenMP.\cite{dagum1998openmp}  
Despite the simplicity of our design, we achieve good parallel scaling, particularly when using a relatively-large $l_\text{max}$ so that there is a substantial amount of computation for each thread, as illustrated in Fig.~\ref{fig:scaling}. The curves show some degree of irregular behavior, suggesting that there might be room for further optimization.
As shown in Table~\ref{tab:cpu-timings}, for large $\lmax$ and numbers of data points, there is a substantial performance gain by using single-precision floating-point arithmetics. However, for smaller amounts of computation there is very small advantage, and in some corner cases ($\lmax=8$, $n_\text{samples}=10,000$, $n_\text{threads}=16$) the single-precision version can be \emph{slower} than that using 64-bit floating-point values.

\begin{table}[bthp]
    \centering
    \renewcommand{\arraystretch}{1.18}
    \begin{tabularx}{\linewidth}{C CC CC}
        \hline\hline
         & \multicolumn{2}{c}{1 OpenMP thread} & \multicolumn{2}{c}{16 OpenMP threads} \\
         & $\tilde{Y}_l^m$ & $\tilde{Y}_l^m, \grad\tilde{Y}_l^m$ & $\tilde{Y}_l^m$ & $\tilde{Y}_l^m, \grad\tilde{Y}_l^m$\\
         \hline
                          &\multicolumn{4}{c}{Single Precision} \\
         \hline
            $l_\text{max}=1$  & 1.08  &  2.00   & 0.572  & 0.680  \\
            $l_\text{max}=2$  & 3.74  &  12.4   & 0.564  & 1.10   \\
            $l_\text{max}=4$  & 9.47  &  35.8   & 0.922  & 2.58   \\
            $l_\text{max}=8$  & 56.4  &  169    & 17.5   & 54.2   \\
            $l_\text{max}=16$ & 240   &  1093   & 28.9   & 126    \\
            $l_\text{max}=32$ & 1128  &  3101   & 109    & 366    \\
         \hline
                  &\multicolumn{4}{c}{Double Precision} \\
         \hline
            $l_\text{max}=1$  &  1.06  &  6.47  &  0.451  &  0.573  \\
            $l_\text{max}=2$  &  4.02  &  15.9  &  0.578  &  1.28   \\
            $l_\text{max}=4$  &  11.6  &  47.3  &  1.00   &  3.73   \\
            $l_\text{max}=8$  &  57.0  &  280   &  5.15   &  21.7   \\
            $l_\text{max}=16$ &  252   &  1602  &  18.3   &  164    \\
            $l_\text{max}=32$ &  1385  &  4651  &  148    &  440    \\
        \hline\hline
    \end{tabularx}
    \caption{CPU-parallel execution time (in ns/point) for computing Cartesian spherical harmonics (or Cartesian spherical harmonics and their derivatives) up to the indicated value of $l_\text{max}$. Timings refer to 1 and 16 OpenMP threads respectively on Intel Xeon Gold 6226R CPU, averaged over 1000 calls for 10~000 points.}
    \label{tab:cpu-timings}
\end{table}

\subsection{CUDA implementation details}

We adapt the implementation to a custom CUDA kernel, that allows efficient execution on GPUs. Similarly to the CPU implementation, we use hard-coded spherical harmonics and derivatives but up to a reduced hard-coded $l_\text{max} =3$ to lower shared memory requirements, and we use the general expression for the remaining terms. 
We parallelize the computation with a two-dimensional thread-block of $16\times8$ threads, using a grid dimension of $(n_\text{samples}/16)$ blocks. 
The first dimension in the thread-block parallelizes over samples, while the first thread in the second dimension is responsible for performing the computation, and the remaining perform coalesced writing of the temporary buffers to global memory. 
We store the intermediary work buffers for the spherical harmonics and derivatives in fast shared memory, which defaults to 48KB for most NVIDIA cards. Memory accesses are designed such that each sample-dimension thread is mapped to shared memory bank contiguously, eliminating possible bank conflicts. The CUDA wrapper automatically re-allocates the necessary amount of shared memory beyond the default 48KB on the first compute call, or attempts to reduce the number of threads launched in the CUDA kernel if the required allocation is too large. As a consequence, for accurate timings we call the forward step once to perform the initialization and then measure timings thereafter. We note that using the recursion~\eqref{eq:iter-q} would require storing temporary values for all $l\le \lmax$, so that the default 48KB shared memory allocation would be filled for $\lmax\approx 8$ when computing derivatives in 64-bit floating-point format.
Using the alternative recursion~\eqref{eq:iter-q-alt} allows us to evaluate one $l$ channel at a time, so values up to $\lmax\approx 30$ can be computed without adjusting the shared memory allocation or reducing the number of sample-dimension threads. For GPUs which support more than 48KB shared memory per block, for example the A100 (164KB), the recursion~\eqref{eq:iter-q-alt} allows us to evaluate values up to $\lmax\approx 82$.

Table~\ref{tab:gpu-timings} shows timings for GPU-accelerated computations on an A100 SXM4 (80GB) card. In general, using single-precision floating-point arithmetic results in half the computation time than double-precision arithmetic. We note that the GPU per-sample timings reduce significantly upon increasing number of samples, as the GPU is not fully saturated for $n_\text{samples} = 10\,000$. For example for $n_\text{samples} = 100\,000$, the per-sample timings reduce by a factor of 1.5 to 4. Although the parallel CPU implementation is faster than the GPU for $l_\text{max} \leq 4$ when using $n_\text{samples}=10\,000$, this trend reverses for higher values of $l_\text{max}$. For example, the 64-bit floating-point computations with $n_\text{samples} = 10\,000$ and $l_\text{max}=16$ show the GPU outperforming the CPU by a factor of 8. 

We note in closing that, even though precise timings depend on the hardware and the maximum angular momentum considered, {\texttt sphericart} is between 10 and 40 times faster than the spherical harmonics implementation in {\texttt e3nn}\cite{e3nn_repo}, a widely used library for equivariant neural networks.
We provide wrappers that use the same conventions as {\texttt{e3nn}}, to simplify integration of our implementation in existing models and to accelerate computational frameworks that are limited by the evaluation of $Y_l^m$.

\begin{table}[bthp]
    \centering
    \renewcommand{\arraystretch}{1.18}
    \begin{tabularx}{\linewidth}{C CC CC}
        \hline\hline
         & \multicolumn{2}{c}{10k points} & \multicolumn{2}{c}{100k points} \\
         & $\tilde{Y}_l^m$ & $\tilde{Y}_l^m, \grad\tilde{Y}_l^m$ & $\tilde{Y}_l^m$ & $\tilde{Y}_l^m, \grad\tilde{Y}_l^m$\\
\hline
        &\multicolumn{4}{c}{Single Precision} \\
         \hline
$l_\text{max}=1$ & 2.1 & 2.4 &  0.3 &  0.4 \\
$l_\text{max}=2$ & 2.2 & 2.4 &  0.3 & 0.4 \\
$l_\text{max}=4$ & 2.4 & 3.1 & 0.5 & 0.8 \\
$l_\text{max}=8$ & 3.1 & 4.8 & 1.0 & 3.3 \\
$l_\text{max}=16$& 5.6 & 15.1 & 2.8 & 10.9 \\
$l_\text{max}=32$& 15.5 & 43.6 & 9.6 & 33.5 \\
         \hline
                  &\multicolumn{4}{c}{Double Precision} \\
         \hline
$l_\text{max}=1$ & 2.2 & 2.6 &  0.4 &  0.6 \\
$l_\text{max}=2$ & 2.3 & 2.9 &  0.4 & 0.8 \\
$l_\text{max}=4$ & 2.7 & 4.0 &  0.7 & 1.7 \\
$l_\text{max}=8$ & 4.2 & 8.5 &  1.7 & 4.6 \\
$l_\text{max}=16$& 9.3 & 22.2 & 5.2 & 16.0 \\
$l_\text{max}=32$& 28.8 & 75.0 & 20.6 & 65.1 \\
         \hline\hline
    \end{tabularx}
    \caption{GPU-parallel execution time (in ns/point) for computing Cartesian spherical harmonics (or Cartesian spherical harmonics and their derivatives) up to the indicated value of $l_\text{max}$. Timings refer to a single A100 SXM4/80GB GPU, and are averaged over 10~000 calls on 10~000 and 100~000 points respectively.}
    \label{tab:gpu-timings}
\end{table}

\section{Conclusions and perspectives}

Spherical harmonics are ubiquitous in the computational sciences, and their efficient calculation has become even more important in light of the widespread adoption of equivariant machine-learning models in chemistry. 
Reformulating the calculation of spherical harmonics in a scaled form that corresponds to real-valued polynomials of the Cartesian coordinates provides simple expressions for their recursive evaluation. Derivatives can be obtained with little overhead, re-using the same factors that enter the definition of the scaled $\tilde{Y}_l^m$, and without numerical instabilities.
The conventional form of the spherical harmonics can be recovered easily by computing $\tilde{Y}_l^m$ at $(x/r,y/r,z/r)$ and applying the corresponding correction to the derivatives. In many applications, this normalization may be unnecessary, as the scaling can be incorporated (explicitly or through regression) into additional radial terms. 
We provide an efficient implementation as a C++ library, complete with a C API, Python and PyTorch bindings, that tackles the most common use case in scientific computing and geometric machine learning, i.e., the evaluation of real-valued spherical harmonics for all angular momentum channels up to a given cutoff $l_\text{max}$ and possibly for many 3D points at once. 
The function call is parallelized over the sample direction, and it uses more efficient hard-coded expressions for low-$l$ terms.  Future efforts will focus on the extension of the software library to different programming languages and frameworks, as well as on further optimization on new hardware platforms and improved parallelism.   

\section*{Data and software availability}

The \emph{sphericart} library can be freely downloaded under the Apache License version 2.0 from the public git repository \url{https://github.com/lab-cosmo/sphericart}. A Python package that provides a convenient class to compute spherical harmonics and derivatives from a \emph{NumPy}\cite{harris2020array} or a \emph{PyTorch}\cite{pasz+19nips} array of 3D coordinates is available on PyPI at \url{https://pypi.org/project/sphericart/}. 
The timing data can be generated using the benchmarking code that is included in the software distribution.

\section*{AUTHOR CONTRIBUTIONS}
{Filippo Bigi} and {Michele Ceriotti} conceived the project and performed analytical derivations. All authors contributed to software development and to the writing of the paper.

\section*{Acknowledgements}

The Authors would like to thank Kevin Kazuki Huguenin-Dumittan for useful discussions, Philip Loche for help with the \texttt{sphericart} library, and Prashanth Kanduri for help with the continuous integration framework.
MC acknowledges funding from the European Research Council (ERC) under the European Union’s Horizon 2020 research and innovation programme (grant agreement No 101001890-FIAMMA). GF acknowledges support from the Platform for Advanced Scientific Computing (PASC). MC and FB acknowledge support from the NCCR MARVEL, funded by the Swiss National Science Foundation (grant number 182892).

\providecommand{\noopsort}[1]{}

\appendix
\section{Recurrence relations for the Cartesian spherical harmonics}

In this Appendix, we derive the Cartesian form of the recurrence relations presented in the main text. Starting from Eq. \ref{eq:real-sh-spherical}, it is sufficient to multiply both sides by $r^l$ and multiply and divide the right-hand side by $r_{xy}^{|m|}$ to obtain
\begin{equation}
    r^l Y_l^m = F_l^{|m|} * r^l r_{xy}^{-|m|} P_l^{|m|} *
    \begin{cases}
      r_{xy}^{|m|} \sin{(|m|\phi)} & \text{if $m < 0$}\\
      r_{xy}^0/\sqrt{2} & \text{if $m = 0$}\\
      r_{xy}^{m} \cos{(m\phi)} & \text{if $m > 0$}
    \end{cases}     
\end{equation}
Then, using the definitions in Eq. \ref{eq:define-q}, Eq. \ref{eq:real-sh-cartesian} follows.

Let us now derive Eq. \ref{eq:iter-q}. These relationships can be obtained as a combination of Eqs. \ref{eq:recur-p} and the definiton of $Q_l^m$ in Eq. 8, with the additional observation that $t = \cos{\theta} = z/r$:
\begin{equation}
    Q_0^0 = r^0 r_{xy}^0 P_0^0 = P_0^0 = 1  
\end{equation}
\begin{multline}
    Q_m^m = r^m r_{xy}^{-m} P_m^m = - \, (2m-1) \, \sqrt{1-t^2} \, r^m r_{xy}^{-m} P_{m-1}^{m-1} = \\ - \, (2m-1) \, \sqrt{1-(z/r)^2} \, (r/r_{xy}) \, Q_{m-1}^{m-1} = - \, (2m-1) \, Q_{m-1}^{m-1}
\end{multline}
\begin{multline}
    Q_m^{m-1} = r^m r_{xy}^{-(m-1)} P_m^{m-1} = \\ (2m-1) \, rt \, r^{-(m-1)} r_{xy}^{-(m-1)} P_{m-1}^{m-1} = (2m-1) \, z \, Q_{m-1}^{m-1}
\end{multline}
\begin{multline}
    Q_l^m = r^l r_{xy}^{-m} P_l^m = \\ r^l r_{xy}^{-m} ((2l-1)\,t\,P_{l-1}^m + (l+m-1)\,P_{l-2}^m)/(l-m) = \\ ((2l-1)\,rt\,r^{l-1} r_{xy}^{-m} P_{l-1}^m + (l+m-1)\, r^2 r^{l-2} r_{xy}^{-m} P_{l-2}^m)/(l-m) = \\ ((2l-1)\,z\,Q_{l-1}^m + (l+m-1)\, r^2 Q_{l-2}^m)/(l-m)
\end{multline}

And, finally, we can derive the recurrence relations for $s_m$ and $c_m$ in the following way:
\begin{multline}
    s_m = r_{xy}^m \sin{m \phi} = r_{xy}^m \sin{((m-1)\phi + \phi)} = \\ r_{xy}^m (\sin{((m-1)\phi)} \cos{\phi} + \cos{((m-1)\phi)} \sin{\phi}) =  \\ s_{m-1} r_{xy} \cos{\phi} + c_{m-1} r_{xy} \sin{\phi} = xs_{m-1} + yc_{m-1}
\end{multline}
\begin{multline}
    c_m = r_{xy}^m \cos{m \phi} = r_{xy}^m \cos{((m-1)\phi + \phi)} = \\ r_{xy}^m (\cos{((m-1)\phi)} \cos{\phi} - \sin{((m-1)\phi)} \sin{\phi}) = \\ c_{m-1} r_{xy} \cos{\phi} - s_{m-1} r_{xy} \sin{\phi} = xc_{m-1} - ys_{m-1}
\end{multline}
where we have used some well-known trigonometric relations and the fact that $\sin{\phi} = y/r_{xy}$, $\cos{\phi} = x/r_{xy}$.

\section{Cartesian derivatives of the scaled spherical harmonics}

In this Appendix, we  prove the derivative formulas presented in the main text, i.e., Eq. \ref{eq:iter-deri}. Let us start from the derivatives of $s_m$ and $c_m$. To this end, we can rewrite Eq. \ref{eq:iter-sc} in matrix form:
\begin{equation}
    \begin{pmatrix} s_m\\c_m \end{pmatrix} = \begin{pmatrix} x&y\\ -y&x \end{pmatrix} \begin{pmatrix} s_{m-1}\\c_{m-1} \end{pmatrix},
\end{equation}
from which it is easy to see that
\begin{equation}
    \begin{pmatrix} s_m\\c_m \end{pmatrix} = \begin{pmatrix} x&y\\ -y&x \end{pmatrix}^m \begin{pmatrix} s_0\\c_0 \end{pmatrix}.
\end{equation}
Differentiation with respect to $x$ yields
\begin{multline}
    \frac{\partial}{\partial x}\begin{pmatrix} s_m\\c_m \end{pmatrix} = m \begin{pmatrix} 1&0\\ 0&1 \end{pmatrix} \begin{pmatrix} x&y\\ -y&x \end{pmatrix}^{m-1} \begin{pmatrix} s_0\\c_0 \end{pmatrix} = \\ m \begin{pmatrix} s_{m-1}\\c_{m-1} \end{pmatrix},
\end{multline}
which proves the $x$ derivatives of $s_m$ and $c_m$ in Eq. \ref{eq:iter-deri}, while differentiation with respect to $y$ gives 
\begin{multline}
    \frac{\partial}{\partial y}\begin{pmatrix} s_m\\c_m \end{pmatrix} = m \begin{pmatrix} 0&1\\ -1&0 \end{pmatrix} \begin{pmatrix} x&y\\ -y&x \end{pmatrix}^{m-1} \begin{pmatrix} s_0\\c_0 \end{pmatrix} = \\ m \begin{pmatrix} 0&1\\ -1&0 \end{pmatrix} \begin{pmatrix} s_{m-1}\\c_{m-1} \end{pmatrix} = m \begin{pmatrix} c_{m-1}\\-s_{m-1} \end{pmatrix},
\end{multline}
which results in the $y$ derivatives of $s_m$ and $c_m$ in Eq. \ref{eq:iter-deri}.

We can now turn to the $\partial Q_l^m / \partial z$ derivative in Eq. \ref{eq:iter-deri}, which we prove by induction over the $l$ variable. The base case ($l = m+1$) can easily be checked for any $m$ by considering the third equality in Eq. \ref{eq:iter-q}, which implies
\begin{equation}
    Q_{m+1}^m = (2m+1) \, z \, Q_m^m.
\end{equation}
Since $Q_m^m$ does not depend on $z$ (see Eq. \ref{eq:iter-deri}), differentiating with respect to $z$ gives 
\begin{equation}
    \frac{\partial Q_{m+1}^m}{\partial z} = (2m+1) \, Q_m^m.
\end{equation}
This is indeed exactly the $\partial Q_l^m / \partial z$ derivative in Eq. \ref{eq:iter-deri} for $l = m+1$. In the induction step, we prove the $l$ case from the $l-1$ and $l-2$ cases, so it is also necessary to prove the $l = m + 2$ case in advance. From the last line in Eq. \ref{eq:iter-q}, we can write 
\begin{equation}
    Q_{m+2}^m = \frac{1}{2} ((2m+3) \, z \, Q_{m+1}^m - (2m+1) \, r^2 \, Q_m^m).
\end{equation}
Differentiation leads to
\begin{multline}
    \frac{\partial{Q_{m+2}^m}}{\partial z} = \frac{1}{2} ((2m+3) \, (Q_{m+1}^m + z \, \frac{\partial{Q_{m+1}^m}}{\partial z}) \\ - (2m+1)(2z \, Q_m^m + r^2 \, \frac{\partial{Q_m^m}}{\partial z})).
\end{multline}
Using $\partial{Q_m^m} / \partial z = 0$ (see Eq. \ref{eq:iter-q}) and $\partial{Q_{m+1}^m} / \partial z = (2m + 1) \, Q_m^m$ (which we have just proved), we obtain
\begin{multline}
    \frac{\partial{Q_{m+2}^m}}{\partial z} = \frac{1}{2} ((2m+3) \, (Q_{m+1}^m + z \, (2m + 1) \, Q_m^m) - \\ (2m+1) \, 2z \, Q_m^m) = (2m + 2) \, Q_{m+1}^m,
\end{multline}
where we have used $(2m+1) \, z \, Q_m^m = Q_{m+1}^m$ from the third equality of Eq. \ref{eq:iter-q} and hidden some elementary algebra. This corresponds to the $\partial Q_l^m / \partial z$ derivative in Eq. \ref{eq:iter-deri} for $l = m + 2$.
For the induction step, we can now assume $\partial Q_{l-1}^m / \partial z = (l+m-1) \, Q_{l-2}^m$ and $\partial Q_{l-2}^m / \partial z = (l+m-2) \, Q_{l-3}^m$. Differentiating the last equality of Eq. \ref{eq:iter-q} with respect to $z$ on both sides results in 
\begin{multline}
    \frac{\partial Q_l^m}{\partial z} = ((2l-1)\,Q_{l-1}^m + (2l-1)\,z\,\frac{\partial Q_{l-1}^m}{\partial z} - \\ (l+m-1)\,2z\,Q_{l-2}^m - (l+m-1)\,r^2\,\frac{\partial Q_{l-2}^m}{\partial z})/(l-m).
\end{multline}
We can now insert the assumptions of the induction step into the right-hand-side expression to obtain
\begin{multline}
    \frac{\partial Q_l^m}{\partial z} = ((2l-1)\,Q_{l-1}^m + (2l-1)\,z\,(l+m-1) \, Q_{l-2}^m - \\ (l+m-1)\,2z\,Q_{l-2}^m - (l+m-1)\,r^2\,(l+m-2) \, Q_{l-3}^m)/(l-m).
\end{multline}
An elementary manipulation of the $z \, Q_{l-2}^m$ terms affords 
\begin{multline}
    \frac{\partial Q_l^m}{\partial z} = ((2l-1)\,Q_{l-1}^m + (l+m-1)(2l-3) \,z\,Q_{l-2}^m - \\ (l+m-1) \, (l+m-2) \, r^2 \, Q_{l-3}^m)/(l-m).
\end{multline}
Finally, application of the last equality of Eq. \ref{eq:iter-q} absorbs the $Q_{l-2}^m$ and $Q_{l-3}^m$ terms into
\begin{equation}
    \frac{\partial Q_l^m}{\partial z} = ((2l-1)\,Q_{l-1}^m + (l+m-1)^2 \, Q_{l-1}^m)/(l-m),
\end{equation}
which simplifies into
\begin{equation}
    \frac{\partial Q_l^m}{\partial z} = (l+m) \, Q_{l-1}^m,
\end{equation}
as required.

To prove the expressions for $\partial Q_l^m / \partial x$ and $\partial Q_l^m / \partial y$ in Eq. \ref{eq:iter-deri}, we first note that, since $Q_l^m$ is formally only a function of $z$ and $r$, any $x$- or $y$-dependence in $Q_l^m$ must come from the involvement of the $r$ variable, so that  
\begin{equation}
    \frac{\partial Q_l^m}{\partial x} = \frac{\partial Q_l^m}{\partial r} \frac{\partial r}{\partial x} = \frac{\partial Q_l^m}{\partial r} \frac{x}{r}
\end{equation}
and
\begin{equation}
    \frac{\partial Q_l^m}{\partial y} = \frac{\partial Q_l^m}{\partial r} \frac{\partial r}{\partial y} = \frac{\partial Q_l^m}{\partial r} \frac{y}{r}.
\end{equation}
Hence, in order to justify the $x$ and $y$ derivatives of $Q_l^m$ in Eq. \ref{eq:iter-deri}, we simply need to prove that 
\begin{equation}\label{eq:r-derivative-equation}
    \frac{\partial Q_l^m}{\partial r} = r \, Q_{l-1}^{m+1}.
\end{equation}
Before doing so, we  borrow Eq. 1 from Ref. \cite{bosch2000computation}. It is important to note how Ref. \cite{bosch2000computation} follows a different convention, and it does not include a $(-1)^m$ factor in the definition of the associated Legendre polynomials. Hence, compared to Ref. \cite{bosch2000computation}, we change the sign of the $P_{l-1}^{m-1}$ term to obtain
\begin{equation}
    P_l^m = - \, (2l-1) \, \sin{\theta} P_{l-1}^{m-1} + P_{l-2}^m,
\end{equation}
which is now consistent with our conventions. Noting that $\sin{\theta} = r_{xy}/r$ and multiplying by $r^l r_{xy}^{-m}$ on both sides gives
\begin{multline}
    r^l r_{xy}^{-m} P_l^m = \\ - \, (2l-1) (r_{xy}/r) \, r^l r_{xy}^{-m} P_{l-1}^{m-1} + r^l r_{xy}^{-m} P_{l-2}^m,
\end{multline}
which, thanks to the definition of $Q_l^m$ in Eq. \ref{eq:define-q}, translates to 
\begin{equation}\label{eq:from-bosch}
    Q_l^m = - \, (2l-1) \, Q_{l-1}^{m-1} + r^2 Q_{l-2}^m.
\end{equation}
Eq. \ref{eq:from-bosch} and the last line of Eq. \ref{eq:iter-q} provide two expressions for $Q_l^m$. Imposing their equality results in
\begin{multline}
    ((2l-1)\,z\,Q_{l-1}^m + (l+m-1)\,r^2\,Q_{l-2}^m)/(l-m) = \\ - \, (2l-1) \, Q_{l-1}^{m-1} + r^2 Q_{l-2}^m.
\end{multline}
This can be rearranged into
\begin{equation}
    z \, Q_{l-1}^m = - \, (l-m) \, Q_{l-1}^{m-1} + r^2 Q_{l-2}^m.
\end{equation}
From this equation, $Q_{l-1}^{m-1}$ can be extracted as 
\begin{equation}\label{eq:final-from-bosch}
    Q_{l-1}^{m-1} = (- \, z \, Q_{l-1}^m + r^2 Q_{l-2}^m)/(l-m).
\end{equation}
Let us now go back to Eq. \ref{eq:r-derivative-equation}. We prove it by induction over the $l$ variable, similar to what we did for the $z$-derivative. Given the $m \leq l$ constraint, the base case reads $\partial Q_{m+1}^{m-1} / \partial r = r \, Q_m^m.$
From the third equality in Eq. \ref{eq:iter-q}, we have $Q_m^{m-1} = (2m-1) \, z \, Q_{m-1}^{m-1}$. Now, using the last line of Eq. \ref{eq:iter-q} with $l = m+1$ and (abusing notation) $m = m-1$, we can express $Q_{m+1}^{m-1}$ as 
\begin{multline}\label{eq:r-derivative-base-step}
    Q_{m+1}^{m-1} = ((2m+1)\,z\,Q_m^{m-1} - (2m-1)\,r^2\,Q_{m-1}^{m-1})/2 = \\ ((2m+1)(2m-1)\, z^2 \, Q_{m-1}^{m-1} - (2m-1)\,r^2\,Q_{m-1}^{m-1})/2 = \\ (2m-1) \, Q_{m-1}^{m-1} \, ((2m+1) \, z^2 - r^2)/2 = \\ - \, Q_m^m \, ((2m+1) \, z^2 - r^2)/2,
\end{multline}
where we have used the second equality of Eq. \ref{eq:iter-q} in the last line. Since $Q_m^m$ does not depend on $r$ (see Eq. \ref{eq:iter-q}), differentiating both sides of Eq. \ref{eq:r-derivative-base-step} with respect to $r$ yields
\begin{equation}
   \frac{\partial Q_{m+1}^{m-1}}{\partial r} = r \, Q_m^m,
\end{equation}
as required. As in the case of the $z$ derivative, the base case also includes a further equality: $\partial Q_{m+2}^{m-1} / \partial r = r \, Q_{m}^{m+1}$. Using the last line of Eq. \ref{eq:iter-q}, $Q_{m+2}^{m-1}$ can be written as 
\begin{equation}
    Q_{m+2}^{m-1} = \frac{1}{3} ((2m+3)\,z\,Q_{m+1}^{m-1} - 2m\,r^2\,Q_{m}^{m-1}).
\end{equation}
partial differentiation of the above with respect to $r$ yields
\begin{multline}
    \frac{\partial Q_{m+2}^{m-1}}{\partial r} = \frac{1}{3} ((2m+3)\,z\,\frac{\partial Q_{m+1}^{m-1}}{\partial r} - 2m \, (2r\,Q_{m}^{m-1} + r^2\,\frac{\partial Q_{m}^{m-1}}{\partial r})).
\end{multline}
However, we have proved $\partial Q_{m+1}^{m-1} / \partial r = r \, Q_m^m$, and we have $\partial Q_{m}^{m-1} / \partial r = 0$ from the first three equalities of Eq. \ref{eq:iter-q}. Hence, 
\begin{multline}
    \frac{\partial Q_{m+2}^{m-1}}{\partial r} = \frac{1}{3} ((2m+3)\,z\,r \, Q_m^m - 4mr\,Q_{m}^{m-1}) = \\ \frac{1}{3} ((2m+3)\,z\,r \, Q_m^m + 4mrz\,Q_m^m) = (2m+1)\,zr\,Q_m^m = r \, Q_{m}^{m+1},
\end{multline}
where we have used the third equality of \ref{eq:iter-q} twice. This concludes the proof of the base cases.

To prove the induction step, we start by differentiating both sides of the last equality of Eq. \ref{eq:iter-q} with respect to $r$. We obtain 
\begin{multline}
    \frac{\partial Q_l^m}{\partial r} = ((2l-1)\,z\frac{\partial Q_{l-1}^m}{\partial r} - 2(l+m-1)\,r\,Q_{l-2}^m - \\ (l+m-1)\,r^2\,\frac{\partial Q_{l-2}^m}{\partial r})/(l-m).
\end{multline}
We can now assume that $\partial Q_{l-1}^m / \partial r = r \, Q_{l-2}^{m+1}$ and $\partial Q_{l-2}^m / \partial r = r \, Q_{l-3}^{m+1}$ to get
\begin{multline}\label{eq:ugly}
    \frac{\partial Q_l^m}{\partial r} = ((2l-1)\,zr \, Q_{l-2}^{m+1} - 2(l+m-1)\,r\,Q_{l-2}^m - \\ (l+m-1)\,r^3 \, Q_{l-3}^{m+1})/(l-m).
\end{multline}
Now, Eq. \ref{eq:final-from-bosch} implies $Q_{l-2}^{m} = (- \, z \, Q_{l-2}^{m+1} + r^2 Q_{l-3}^{m+1})/(l-m-2)$. Substituting this into Eq. \ref{eq:ugly} leads to
\begin{multline}
    \frac{\partial Q_l^m}{\partial r} = ((2l-1)\,zr \, Q_{l-2}^{m+1} - \\ \frac{2(l+m-1)}{l-m-2}\,(- \, zr \, Q_{l-2}^{m+1} + r^3 Q_{l-3}^{m+1}) - \\ (l+m-1)\,r^3 \, Q_{l-3}^{m+1})/(l-m).
\end{multline}
Adding like terms (i.e., those in $zr\,Q_{l-2}^{m+1}$ and those in $r^3\,Q_{l-3}^{m+1}$) results in
\begin{multline}
    \frac{\partial Q_l^m}{\partial r} = \frac{1}{(l-m)(l-m-2)}((2l-3)(l-m)\,zr \, Q_{l-2}^{m+1} - \\ (l+m-1)(l-m) \, r^3 \, Q_{l-3}^{m+1}) = \\ \frac{1}{l-m-2}((2l-3)\,zr \, Q_{l-2}^{m+1} - (l+m-1) \, r^3 \, Q_{l-3}^{m+1}) = \\ r \frac{1}{l-m-2}((2l-3)\,z \, Q_{l-2}^{m+1} - (l+m-1) \, r^2 \, Q_{l-3}^{m+1}).
\end{multline}
However, by considering the last line of Eq. \ref{eq:iter-q} with $l$ decreased by one and m increased by one, the last expression can be simplified into
\begin{equation}
    \frac{\partial Q_l^m}{\partial r} = r \, Q_{l-1}^{m+1},
\end{equation}
which concludes the proof.

\section{Considerations on the prefactor}\label{app:Flm}

The prefactors $F_l^m$ contain a ratio of factorials that can lead to numerical instabilities in a na\"ive implementation. It is however easy to see that one can compute them iteratively as 
\begin{equation}
F_l^0 = \sqrt{\frac{2l+1}{2\pi}}, \quad F_l^m = -\frac{F_l^{m-1}}{\sqrt{(l+m)(l+1-m)}}.
\end{equation}
It is also possible to incorporate the prefactors in the definition of the modified associated Legendre polynomials, defining $\tilde{Q}_l^m=F_l^m {Q}_l^m$. This simplifies somehow the construction of $\tilde{Y}_l^m$ and avoids possible instabilities connected with the fact that $F_l^m$ become very small for large $m\approx l$, at the price of complicating slightly the expressions for the recursion and  derivatives of $Q_l^m$, e.g.
\begin{equation}
\frac{\partial \tilde{Q}_l^m}{\partial z} = \sqrt{l^2-m^2} \,\tilde{Q}_{l-1}^m.
\end{equation}


\begin{thebibliography}{40}%
\makeatletter
\providecommand \@ifxundefined [1]{%
 \@ifx{#1\undefined}
}%
\providecommand \@ifnum [1]{%
 \ifnum #1\expandafter \@firstoftwo
 \else \expandafter \@secondoftwo
 \fi
}%
\providecommand \@ifx [1]{%
 \ifx #1\expandafter \@firstoftwo
 \else \expandafter \@secondoftwo
 \fi
}%
\providecommand \natexlab [1]{#1}%
\providecommand \enquote  [1]{``#1''}%
\providecommand \bibnamefont  [1]{#1}%
\providecommand \bibfnamefont [1]{#1}%
\providecommand \citenamefont [1]{#1}%
\providecommand \href@noop [0]{\@secondoftwo}%
\providecommand \href [0]{\begingroup \@sanitize@url \@href}%
\providecommand \@href[1]{\@@startlink{#1}\@@href}%
\providecommand \@@href[1]{\endgroup#1\@@endlink}%
\providecommand \@sanitize@url [0]{\catcode `\\12\catcode `\$12\catcode
  `\&12\catcode `\#12\catcode `\^12\catcode `\_12\catcode `\%12\relax}%
\providecommand \@@startlink[1]{}%
\providecommand \@@endlink[0]{}%
\providecommand \url  [0]{\begingroup\@sanitize@url \@url }%
\providecommand \@url [1]{\endgroup\@href {#1}{\urlprefix }}%
\providecommand \urlprefix  [0]{URL }%
\providecommand \Eprint [0]{\href }%
\providecommand \doibase [0]{http://dx.doi.org/}%
\providecommand \selectlanguage [0]{\@gobble}%
\providecommand \bibinfo  [0]{\@secondoftwo}%
\providecommand \bibfield  [0]{\@secondoftwo}%
\providecommand \translation [1]{[#1]}%
\providecommand \BibitemOpen [0]{}%
\providecommand \bibitemStop [0]{}%
\providecommand \bibitemNoStop [0]{.\EOS\space}%
\providecommand \EOS [0]{\spacefactor3000\relax}%
\providecommand \BibitemShut  [1]{\csname bibitem#1\endcsname}%
\let\auto@bib@innerbib\@empty
\bibitem [{\citenamefont {Müller}(1966)}]{clausmller_1966_spherical}%
  \BibitemOpen
  \bibfield  {author} {\bibinfo {author} {\bibfnamefont {C.}~\bibnamefont
  {Müller}},\ }\href@noop {} {\emph {\bibinfo {title} {Spherical harmonics}}}\
  (\bibinfo  {publisher} {Springer},\ \bibinfo {year} {1966})\BibitemShut
  {NoStop}%
\bibitem [{\citenamefont {Steiner}(1963)}]{steiner1963charge}%
  \BibitemOpen
  \bibfield  {author} {\bibinfo {author} {\bibfnamefont {E.}~\bibnamefont
  {Steiner}},\ }\href@noop {} {\bibfield  {journal} {\bibinfo  {journal} {The
  Journal of Chemical Physics}\ }\textbf {\bibinfo {volume} {39}},\ \bibinfo
  {pages} {2365} (\bibinfo {year} {1963})}\BibitemShut {NoStop}%
\bibitem [{\citenamefont {Rexer}\ \emph {et~al.}(2016)\citenamefont {Rexer},
  \citenamefont {Hirt}, \citenamefont {Claessens},\ and\ \citenamefont
  {Tenzer}}]{rexer_2016_layerbased}%
  \BibitemOpen
  \bibfield  {author} {\bibinfo {author} {\bibfnamefont {M.}~\bibnamefont
  {Rexer}}, \bibinfo {author} {\bibfnamefont {C.}~\bibnamefont {Hirt}},
  \bibinfo {author} {\bibfnamefont {S.}~\bibnamefont {Claessens}}, \ and\
  \bibinfo {author} {\bibfnamefont {R.}~\bibnamefont {Tenzer}},\ }\href
  {\doibase 10.1007/s10712-016-9382-2} {\bibfield  {journal} {\bibinfo
  {journal} {Surveys in Geophysics}\ }\textbf {\bibinfo {volume} {37}},\
  \bibinfo {pages} {1035} (\bibinfo {year} {2016})}\BibitemShut {NoStop}%
\bibitem [{\citenamefont {Knaack}\ and\ \citenamefont
  {Stenflo}(2005)}]{knaack_2005_spherical}%
  \BibitemOpen
  \bibfield  {author} {\bibinfo {author} {\bibfnamefont {R.}~\bibnamefont
  {Knaack}}\ and\ \bibinfo {author} {\bibfnamefont {J.~O.}\ \bibnamefont
  {Stenflo}},\ }\href {\doibase 10.1051/0004-6361:20052765} {\bibfield
  {journal} {\bibinfo  {journal} {Astronomy and Astrophysics}\ }\textbf
  {\bibinfo {volume} {438}},\ \bibinfo {pages} {349} (\bibinfo {year}
  {2005})}\BibitemShut {NoStop}%
\bibitem [{\citenamefont {Morschhauser}\ \emph {et~al.}(2014)\citenamefont
  {Morschhauser}, \citenamefont {Lesur},\ and\ \citenamefont
  {Grott}}]{morschhauser_2014_a}%
  \BibitemOpen
  \bibfield  {author} {\bibinfo {author} {\bibfnamefont {A.}~\bibnamefont
  {Morschhauser}}, \bibinfo {author} {\bibfnamefont {V.}~\bibnamefont {Lesur}},
  \ and\ \bibinfo {author} {\bibfnamefont {M.}~\bibnamefont {Grott}},\ }\href
  {\doibase 10.1002/2013je004555} {\bibfield  {journal} {\bibinfo  {journal}
  {Journal of Geophysical Research: Planets}\ }\textbf {\bibinfo {volume}
  {119}},\ \bibinfo {pages} {1162} (\bibinfo {year} {2014})}\BibitemShut
  {NoStop}%
\bibitem [{\citenamefont {Evans}(1998)}]{evans_1998_the}%
  \BibitemOpen
  \bibfield  {author} {\bibinfo {author} {\bibfnamefont {K.~F.}\ \bibnamefont
  {Evans}},\ }\href {\doibase 10.1175/1520-0469(1998)055<0429:tshdom>2.0.co;2}
  {\bibfield  {journal} {\bibinfo  {journal} {Journal of the Atmospheric
  Sciences}\ }\textbf {\bibinfo {volume} {55}},\ \bibinfo {pages} {429}
  (\bibinfo {year} {1998})}\BibitemShut {NoStop}%
\bibitem [{\citenamefont {poletti}(2005)}]{poletti2005three-dimensional}%
  \BibitemOpen
  \bibfield  {author} {\bibinfo {author} {\bibfnamefont {m.~a.}\ \bibnamefont
  {poletti}},\ }\href@noop {} {\bibfield  {journal} {\bibinfo  {journal}
  {journal of the audio engineering society}\ }\textbf {\bibinfo {volume}
  {53}},\ \bibinfo {pages} {1004} (\bibinfo {year} {2005})}\BibitemShut
  {NoStop}%
\bibitem [{\citenamefont {Max}\ and\ \citenamefont
  {Getzoff}(1988)}]{max1988spherical}%
  \BibitemOpen
  \bibfield  {author} {\bibinfo {author} {\bibfnamefont {N.~L.}\ \bibnamefont
  {Max}}\ and\ \bibinfo {author} {\bibfnamefont {E.~D.}\ \bibnamefont
  {Getzoff}},\ }\href@noop {} {\bibfield  {journal} {\bibinfo  {journal} {IEEE
  Computer Graphics and Applications}\ }\textbf {\bibinfo {volume} {8}},\
  \bibinfo {pages} {42} (\bibinfo {year} {1988})}\BibitemShut {NoStop}%
\bibitem [{\citenamefont {Sloan}(2008)}]{sloan2008stupid}%
  \BibitemOpen
  \bibfield  {author} {\bibinfo {author} {\bibfnamefont {P.-P.}\ \bibnamefont
  {Sloan}},\ }in\ \href@noop {} {\emph {\bibinfo {booktitle} {Game developers
  conference}}},\ Vol.~\bibinfo {volume} {9}\ (\bibinfo {year} {2008})\
  p.~\bibinfo {pages} {42}\BibitemShut {NoStop}%
\bibitem [{\citenamefont {Schlegel}\ and\ \citenamefont
  {Frisch}(1995)}]{schlegel1995transformation}%
  \BibitemOpen
  \bibfield  {author} {\bibinfo {author} {\bibfnamefont {H.~B.}\ \bibnamefont
  {Schlegel}}\ and\ \bibinfo {author} {\bibfnamefont {M.~J.}\ \bibnamefont
  {Frisch}},\ }\href@noop {} {\bibfield  {journal} {\bibinfo  {journal}
  {International Journal of Quantum Chemistry}\ }\textbf {\bibinfo {volume}
  {54}},\ \bibinfo {pages} {83} (\bibinfo {year} {1995})}\BibitemShut {NoStop}%
\bibitem [{\citenamefont {Varganov}\ \emph {et~al.}(2008)\citenamefont
  {Varganov}, \citenamefont {Gilbert}, \citenamefont {Deplazes},\ and\
  \citenamefont {Gill}}]{varganov2008resolutions}%
  \BibitemOpen
  \bibfield  {author} {\bibinfo {author} {\bibfnamefont {S.~A.}\ \bibnamefont
  {Varganov}}, \bibinfo {author} {\bibfnamefont {A.~T.}\ \bibnamefont
  {Gilbert}}, \bibinfo {author} {\bibfnamefont {E.}~\bibnamefont {Deplazes}}, \
  and\ \bibinfo {author} {\bibfnamefont {P.~M.}\ \bibnamefont {Gill}},\
  }\href@noop {} {\bibfield  {journal} {\bibinfo  {journal} {The Journal of
  chemical physics}\ }\textbf {\bibinfo {volume} {128}},\ \bibinfo {pages}
  {201104} (\bibinfo {year} {2008})}\BibitemShut {NoStop}%
\bibitem [{\citenamefont {Gill}\ and\ \citenamefont
  {Gilbert}(2009)}]{gill2009resolutions}%
  \BibitemOpen
  \bibfield  {author} {\bibinfo {author} {\bibfnamefont {P.~M.}\ \bibnamefont
  {Gill}}\ and\ \bibinfo {author} {\bibfnamefont {A.~T.}\ \bibnamefont
  {Gilbert}},\ }\href@noop {} {\bibfield  {journal} {\bibinfo  {journal}
  {Chemical Physics}\ }\textbf {\bibinfo {volume} {356}},\ \bibinfo {pages}
  {86} (\bibinfo {year} {2009})}\BibitemShut {NoStop}%
\bibitem [{\citenamefont {Maintz}\ \emph {et~al.}(2016)\citenamefont {Maintz},
  \citenamefont {Esser},\ and\ \citenamefont
  {Dronskowski}}]{maintz2016efficient}%
  \BibitemOpen
  \bibfield  {author} {\bibinfo {author} {\bibfnamefont {S.}~\bibnamefont
  {Maintz}}, \bibinfo {author} {\bibfnamefont {M.}~\bibnamefont {Esser}}, \
  and\ \bibinfo {author} {\bibfnamefont {R.}~\bibnamefont {Dronskowski}},\
  }\href@noop {} {\bibfield  {journal} {\bibinfo  {journal} {Acta Physica
  Polonica B}\ }\textbf {\bibinfo {volume} {47}} (\bibinfo {year}
  {2016})}\BibitemShut {NoStop}%
\bibitem [{\citenamefont {P{\'e}rez-Jord{\'a}}\ and\ \citenamefont
  {Yang}(1996)}]{pere-yang96jcp}%
  \BibitemOpen
  \bibfield  {author} {\bibinfo {author} {\bibfnamefont {J.~M.}\ \bibnamefont
  {P{\'e}rez-Jord{\'a}}}\ and\ \bibinfo {author} {\bibfnamefont
  {W.}~\bibnamefont {Yang}},\ }\href {\doibase 10.1063/1.471517} {\bibfield
  {journal} {\bibinfo  {journal} {The Journal of Chemical Physics}\ }\textbf
  {\bibinfo {volume} {104}},\ \bibinfo {pages} {8003} (\bibinfo {year}
  {1996})}\BibitemShut {NoStop}%
\bibitem [{\citenamefont {Choi}\ \emph {et~al.}(1999)\citenamefont {Choi},
  \citenamefont {Ivanic}, \citenamefont {Gordon},\ and\ \citenamefont
  {Ruedenberg}}]{choi+99jcp}%
  \BibitemOpen
  \bibfield  {author} {\bibinfo {author} {\bibfnamefont {C.~H.}\ \bibnamefont
  {Choi}}, \bibinfo {author} {\bibfnamefont {J.}~\bibnamefont {Ivanic}},
  \bibinfo {author} {\bibfnamefont {M.~S.}\ \bibnamefont {Gordon}}, \ and\
  \bibinfo {author} {\bibfnamefont {K.}~\bibnamefont {Ruedenberg}},\ }\href
  {\doibase 10.1063/1.480229} {\bibfield  {journal} {\bibinfo  {journal} {The
  Journal of Chemical Physics}\ }\textbf {\bibinfo {volume} {111}},\ \bibinfo
  {pages} {8825} (\bibinfo {year} {1999})}\BibitemShut {NoStop}%
\bibitem [{\citenamefont {Ding}\ \emph {et~al.}(2017)\citenamefont {Ding},
  \citenamefont {Levesque}, \citenamefont {Borgis},\ and\ \citenamefont
  {Belloni}}]{ding+17jcp}%
  \BibitemOpen
  \bibfield  {author} {\bibinfo {author} {\bibfnamefont {L.}~\bibnamefont
  {Ding}}, \bibinfo {author} {\bibfnamefont {M.}~\bibnamefont {Levesque}},
  \bibinfo {author} {\bibfnamefont {D.}~\bibnamefont {Borgis}}, \ and\ \bibinfo
  {author} {\bibfnamefont {L.}~\bibnamefont {Belloni}},\ }\href {\doibase
  10.1063/1.4994281} {\bibfield  {journal} {\bibinfo  {journal} {The Journal of
  Chemical Physics}\ }\textbf {\bibinfo {volume} {147}},\ \bibinfo {pages}
  {094107} (\bibinfo {year} {2017})}\BibitemShut {NoStop}%
\bibitem [{\citenamefont {Zotkin}\ \emph {et~al.}(2009)\citenamefont {Zotkin},
  \citenamefont {Duraiswami},\ and\ \citenamefont
  {Gumerov}}]{zotkin2009regularized}%
  \BibitemOpen
  \bibfield  {author} {\bibinfo {author} {\bibfnamefont {D.~N.}\ \bibnamefont
  {Zotkin}}, \bibinfo {author} {\bibfnamefont {R.}~\bibnamefont {Duraiswami}},
  \ and\ \bibinfo {author} {\bibfnamefont {N.~A.}\ \bibnamefont {Gumerov}},\
  }in\ \href@noop {} {\emph {\bibinfo {booktitle} {2009 IEEE Workshop on
  Applications of Signal Processing to Audio and Acoustics}}}\ (\bibinfo
  {organization} {IEEE},\ \bibinfo {year} {2009})\ pp.\ \bibinfo {pages}
  {257--260}\BibitemShut {NoStop}%
\bibitem [{\citenamefont {Li}\ \emph {et~al.}(2011)\citenamefont {Li},
  \citenamefont {Yan}, \citenamefont {Ma},\ and\ \citenamefont
  {Hou}}]{li2011spherical}%
  \BibitemOpen
  \bibfield  {author} {\bibinfo {author} {\bibfnamefont {X.}~\bibnamefont
  {Li}}, \bibinfo {author} {\bibfnamefont {S.}~\bibnamefont {Yan}}, \bibinfo
  {author} {\bibfnamefont {X.}~\bibnamefont {Ma}}, \ and\ \bibinfo {author}
  {\bibfnamefont {C.}~\bibnamefont {Hou}},\ }\href@noop {} {\bibfield
  {journal} {\bibinfo  {journal} {Applied Acoustics}\ }\textbf {\bibinfo
  {volume} {72}},\ \bibinfo {pages} {646} (\bibinfo {year} {2011})}\BibitemShut
  {NoStop}%
\bibitem [{\citenamefont {Cohen}\ and\ \citenamefont
  {Welling}(2016)}]{cohe-well16icml}%
  \BibitemOpen
  \bibfield  {author} {\bibinfo {author} {\bibfnamefont {T.}~\bibnamefont
  {Cohen}}\ and\ \bibinfo {author} {\bibfnamefont {M.}~\bibnamefont
  {Welling}},\ }in\ \href@noop {} {\emph {\bibinfo {booktitle} {Int. {{Conf}}.
  {{Mach}}. {{Learn}}.}}}\ (\bibinfo {organization} {{PMLR}},\ \bibinfo {year}
  {2016})\ pp.\ \bibinfo {pages} {2990--2999}\BibitemShut {NoStop}%
\bibitem [{\citenamefont {Bronstein}\ \emph {et~al.}(2021)\citenamefont
  {Bronstein}, \citenamefont {Bruna}, \citenamefont {Cohen},\ and\
  \citenamefont {Veli{\v c}kovi{\'c}}}]{bron+21arxiv}%
  \BibitemOpen
  \bibfield  {author} {\bibinfo {author} {\bibfnamefont {M.~M.}\ \bibnamefont
  {Bronstein}}, \bibinfo {author} {\bibfnamefont {J.}~\bibnamefont {Bruna}},
  \bibinfo {author} {\bibfnamefont {T.}~\bibnamefont {Cohen}}, \ and\ \bibinfo
  {author} {\bibfnamefont {P.}~\bibnamefont {Veli{\v c}kovi{\'c}}},\ }\href
  {http://arxiv.org/abs/2104.13478v2} {\bibfield  {journal} {\bibinfo
  {journal} {arxiv:2104.13478}\ } (\bibinfo {year} {2021})}\BibitemShut
  {NoStop}%
\bibitem [{\citenamefont {Thomas}\ \emph {et~al.}(2018)\citenamefont {Thomas},
  \citenamefont {Smidt}, \citenamefont {Kearnes}, \citenamefont {Yang},
  \citenamefont {Li}, \citenamefont {Kohlhoff},\ and\ \citenamefont
  {Riley}}]{thom+18arxiv}%
  \BibitemOpen
  \bibfield  {author} {\bibinfo {author} {\bibfnamefont {N.}~\bibnamefont
  {Thomas}}, \bibinfo {author} {\bibfnamefont {T.}~\bibnamefont {Smidt}},
  \bibinfo {author} {\bibfnamefont {S.}~\bibnamefont {Kearnes}}, \bibinfo
  {author} {\bibfnamefont {L.}~\bibnamefont {Yang}}, \bibinfo {author}
  {\bibfnamefont {L.}~\bibnamefont {Li}}, \bibinfo {author} {\bibfnamefont
  {K.}~\bibnamefont {Kohlhoff}}, \ and\ \bibinfo {author} {\bibfnamefont
  {P.}~\bibnamefont {Riley}},\ }\href {http://arxiv.org/abs/1802.08219v3}
  {\bibfield  {journal} {\bibinfo  {journal} {arxiv:1802.08219}\ } (\bibinfo
  {year} {2018})}\BibitemShut {NoStop}%
\bibitem [{\citenamefont {Anderson}\ \emph {et~al.}(2019)\citenamefont
  {Anderson}, \citenamefont {Hy},\ and\ \citenamefont {Kondor}}]{ande+19nips}%
  \BibitemOpen
  \bibfield  {author} {\bibinfo {author} {\bibfnamefont {B.}~\bibnamefont
  {Anderson}}, \bibinfo {author} {\bibfnamefont {T.~S.}\ \bibnamefont {Hy}}, \
  and\ \bibinfo {author} {\bibfnamefont {R.}~\bibnamefont {Kondor}},\ }in\
  \href@noop {} {\emph {\bibinfo {booktitle} {{{NeurIPS}}}}}\ (\bibinfo {year}
  {2019})\ p.~\bibinfo {pages} {10}\BibitemShut {NoStop}%
\bibitem [{\citenamefont {Klicpera}\ \emph {et~al.}(2021)\citenamefont
  {Klicpera}, \citenamefont {Becker},\ and\ \citenamefont
  {G{\"u}nnemann}}]{klic+21arxiv}%
  \BibitemOpen
  \bibfield  {author} {\bibinfo {author} {\bibfnamefont {J.}~\bibnamefont
  {Klicpera}}, \bibinfo {author} {\bibfnamefont {F.}~\bibnamefont {Becker}}, \
  and\ \bibinfo {author} {\bibfnamefont {S.}~\bibnamefont {G{\"u}nnemann}},\
  }\href {http://arxiv.org/abs/2106.08903v8} {\bibfield  {journal} {\bibinfo
  {journal} {arxiv:2106.08903}\ } (\bibinfo {year} {2021})}\BibitemShut
  {NoStop}%
\bibitem [{\citenamefont {Geiger}\ and\ \citenamefont
  {Smidt}(2022)}]{geiger2022e3nn}%
  \BibitemOpen
  \bibfield  {author} {\bibinfo {author} {\bibfnamefont {M.}~\bibnamefont
  {Geiger}}\ and\ \bibinfo {author} {\bibfnamefont {T.}~\bibnamefont {Smidt}},\
  }\href@noop {} {\bibfield  {journal} {\bibinfo  {journal} {arXiv preprint
  arXiv:2207.09453}\ } (\bibinfo {year} {2022})}\BibitemShut {NoStop}%
\bibitem [{\citenamefont {Bart{\'o}k}\ \emph {et~al.}(2013)\citenamefont
  {Bart{\'o}k}, \citenamefont {Kondor},\ and\ \citenamefont
  {Cs{\'a}nyi}}]{bart+13prb}%
  \BibitemOpen
  \bibfield  {author} {\bibinfo {author} {\bibfnamefont {A.~P.}\ \bibnamefont
  {Bart{\'o}k}}, \bibinfo {author} {\bibfnamefont {R.}~\bibnamefont {Kondor}},
  \ and\ \bibinfo {author} {\bibfnamefont {G.}~\bibnamefont {Cs{\'a}nyi}},\
  }\href {\doibase 10.1103/PhysRevB.87.184115} {\bibfield  {journal} {\bibinfo
  {journal} {Phys. Rev. B}\ }\textbf {\bibinfo {volume} {87}},\ \bibinfo
  {pages} {184115} (\bibinfo {year} {2013})}\BibitemShut {NoStop}%
\bibitem [{\citenamefont {Thompson}\ \emph {et~al.}(2015)\citenamefont
  {Thompson}, \citenamefont {Swiler}, \citenamefont {Trott}, \citenamefont
  {Foiles},\ and\ \citenamefont {Tucker}}]{thom+15jcp}%
  \BibitemOpen
  \bibfield  {author} {\bibinfo {author} {\bibfnamefont {A.}~\bibnamefont
  {Thompson}}, \bibinfo {author} {\bibfnamefont {L.}~\bibnamefont {Swiler}},
  \bibinfo {author} {\bibfnamefont {C.}~\bibnamefont {Trott}}, \bibinfo
  {author} {\bibfnamefont {S.}~\bibnamefont {Foiles}}, \ and\ \bibinfo {author}
  {\bibfnamefont {G.}~\bibnamefont {Tucker}},\ }\href {\doibase
  10.1016/j.jcp.2014.12.018} {\bibfield  {journal} {\bibinfo  {journal}
  {Journal of Computational Physics}\ }\textbf {\bibinfo {volume} {285}},\
  \bibinfo {pages} {316} (\bibinfo {year} {2015})}\BibitemShut {NoStop}%
\bibitem [{\citenamefont {Willatt}\ \emph {et~al.}(2019)\citenamefont
  {Willatt}, \citenamefont {Musil},\ and\ \citenamefont
  {Ceriotti}}]{will+19jcp}%
  \BibitemOpen
  \bibfield  {author} {\bibinfo {author} {\bibfnamefont {M.~J.}\ \bibnamefont
  {Willatt}}, \bibinfo {author} {\bibfnamefont {F.}~\bibnamefont {Musil}}, \
  and\ \bibinfo {author} {\bibfnamefont {M.}~\bibnamefont {Ceriotti}},\ }\href
  {\doibase 10.1063/1.5090481} {\bibfield  {journal} {\bibinfo  {journal} {J.
  Chem. Phys.}\ }\textbf {\bibinfo {volume} {150}},\ \bibinfo {pages} {154110}
  (\bibinfo {year} {2019})}\BibitemShut {NoStop}%
\bibitem [{\citenamefont {Drautz}(2019)}]{drau19prb}%
  \BibitemOpen
  \bibfield  {author} {\bibinfo {author} {\bibfnamefont {R.}~\bibnamefont
  {Drautz}},\ }\href {\doibase 10.1103/PhysRevB.99.014104} {\bibfield
  {journal} {\bibinfo  {journal} {Phys. Rev. B}\ }\textbf {\bibinfo {volume}
  {99}},\ \bibinfo {pages} {014104} (\bibinfo {year} {2019})}\BibitemShut
  {NoStop}%
\bibitem [{\citenamefont {Musil}\ \emph {et~al.}(2021)\citenamefont {Musil},
  \citenamefont {Grisafi}, \citenamefont {Bart{\'o}k}, \citenamefont {Ortner},
  \citenamefont {Cs{\'a}nyi},\ and\ \citenamefont {Ceriotti}}]{musi+21cr}%
  \BibitemOpen
  \bibfield  {author} {\bibinfo {author} {\bibfnamefont {F.}~\bibnamefont
  {Musil}}, \bibinfo {author} {\bibfnamefont {A.}~\bibnamefont {Grisafi}},
  \bibinfo {author} {\bibfnamefont {A.~P.}\ \bibnamefont {Bart{\'o}k}},
  \bibinfo {author} {\bibfnamefont {C.}~\bibnamefont {Ortner}}, \bibinfo
  {author} {\bibfnamefont {G.}~\bibnamefont {Cs{\'a}nyi}}, \ and\ \bibinfo
  {author} {\bibfnamefont {M.}~\bibnamefont {Ceriotti}},\ }\href {\doibase
  10.1021/acs.chemrev.1c00021} {\bibfield  {journal} {\bibinfo  {journal}
  {Chem. Rev.}\ }\textbf {\bibinfo {volume} {121}},\ \bibinfo {pages} {9759}
  (\bibinfo {year} {2021})}\BibitemShut {NoStop}%
\bibitem [{\citenamefont {Christensen}\ and\ \citenamefont
  {Von~Lilienfeld}(2020)}]{christensen2020role}%
  \BibitemOpen
  \bibfield  {author} {\bibinfo {author} {\bibfnamefont {A.~S.}\ \bibnamefont
  {Christensen}}\ and\ \bibinfo {author} {\bibfnamefont {O.~A.}\ \bibnamefont
  {Von~Lilienfeld}},\ }\href@noop {} {\bibfield  {journal} {\bibinfo  {journal}
  {Machine Learning: Science and Technology}\ }\textbf {\bibinfo {volume}
  {1}},\ \bibinfo {pages} {045018} (\bibinfo {year} {2020})}\BibitemShut
  {NoStop}%
\bibitem [{\citenamefont {Sch{\"u}tt}\ \emph {et~al.}(2021)\citenamefont
  {Sch{\"u}tt}, \citenamefont {Unke},\ and\ \citenamefont
  {Gastegger}}]{schu+21icml}%
  \BibitemOpen
  \bibfield  {author} {\bibinfo {author} {\bibfnamefont {K.}~\bibnamefont
  {Sch{\"u}tt}}, \bibinfo {author} {\bibfnamefont {O.}~\bibnamefont {Unke}}, \
  and\ \bibinfo {author} {\bibfnamefont {M.}~\bibnamefont {Gastegger}},\ }in\
  \href@noop {} {\emph {\bibinfo {booktitle} {Int. {{Conf}}. {{Mach}}.
  {{Learn}}.}}}\ (\bibinfo {organization} {{PMLR}},\ \bibinfo {year} {2021})\
  pp.\ \bibinfo {pages} {9377--9388}\BibitemShut {NoStop}%
\bibitem [{\citenamefont {Press}(2007)}]{press07book}%
  \BibitemOpen
  \bibfield  {author} {\bibinfo {author} {\bibfnamefont {W.~H.}\ \bibnamefont
  {Press}},\ }\href@noop {} {\emph {\bibinfo {title} {Numerical {{Recipes}}:
  {{The}} Art of Scientific Computing}}}\ (\bibinfo  {publisher} {{Cambridge
  University Press}},\ \bibinfo {year} {2007})\BibitemShut {NoStop}%
\bibitem [{\citenamefont {Sloan}(2013)}]{sloa13jcgt}%
  \BibitemOpen
  \bibfield  {author} {\bibinfo {author} {\bibfnamefont {P.-P.}\ \bibnamefont
  {Sloan}},\ }\href {http://jcgt.org/published/0002/02/06/} {\bibfield
  {journal} {\bibinfo  {journal} {J. Comput. Graph. Tech. JCGT}\ }\textbf
  {\bibinfo {volume} {2}},\ \bibinfo {pages} {84} (\bibinfo {year}
  {2013})}\BibitemShut {NoStop}%
\bibitem [{\citenamefont {Drautz}(2020)}]{drau20prb}%
  \BibitemOpen
  \bibfield  {author} {\bibinfo {author} {\bibfnamefont {R.}~\bibnamefont
  {Drautz}},\ }\href {\doibase 10.1103/PhysRevB.102.024104} {\bibfield
  {journal} {\bibinfo  {journal} {Phys. Rev. B}\ }\textbf {\bibinfo {volume}
  {102}},\ \bibinfo {pages} {024104} (\bibinfo {year} {2020})}\BibitemShut
  {NoStop}%
\bibitem [{\citenamefont {Alken}(2022)}]{alken2022gsl}%
  \BibitemOpen
  \bibfield  {author} {\bibinfo {author} {\bibfnamefont {P.}~\bibnamefont
  {Alken}},\ }\href@noop {} {\  (\bibinfo {year} {2022})}\BibitemShut {NoStop}%
\bibitem [{\citenamefont {Paszke}\ \emph {et~al.}(2019)\citenamefont {Paszke},
  \citenamefont {Gross}, \citenamefont {Massa}, \citenamefont {Lerer},
  \citenamefont {Bradbury}, \citenamefont {Chanan}, \citenamefont {Killeen},
  \citenamefont {Lin}, \citenamefont {Gimelshein}, \citenamefont {Antiga},
  \citenamefont {Desmaison}, \citenamefont {Kopf}, \citenamefont {Yang},
  \citenamefont {DeVito}, \citenamefont {Raison}, \citenamefont {Tejani},
  \citenamefont {Chilamkurthy}, \citenamefont {Steiner}, \citenamefont {Fang},
  \citenamefont {Bai},\ and\ \citenamefont {Chintala}}]{pasz+19nips}%
  \BibitemOpen
  \bibfield  {author} {\bibinfo {author} {\bibfnamefont {A.}~\bibnamefont
  {Paszke}}, \bibinfo {author} {\bibfnamefont {S.}~\bibnamefont {Gross}},
  \bibinfo {author} {\bibfnamefont {F.}~\bibnamefont {Massa}}, \bibinfo
  {author} {\bibfnamefont {A.}~\bibnamefont {Lerer}}, \bibinfo {author}
  {\bibfnamefont {J.}~\bibnamefont {Bradbury}}, \bibinfo {author}
  {\bibfnamefont {G.}~\bibnamefont {Chanan}}, \bibinfo {author} {\bibfnamefont
  {T.}~\bibnamefont {Killeen}}, \bibinfo {author} {\bibfnamefont
  {Z.}~\bibnamefont {Lin}}, \bibinfo {author} {\bibfnamefont {N.}~\bibnamefont
  {Gimelshein}}, \bibinfo {author} {\bibfnamefont {L.}~\bibnamefont {Antiga}},
  \bibinfo {author} {\bibfnamefont {A.}~\bibnamefont {Desmaison}}, \bibinfo
  {author} {\bibfnamefont {A.}~\bibnamefont {Kopf}}, \bibinfo {author}
  {\bibfnamefont {E.}~\bibnamefont {Yang}}, \bibinfo {author} {\bibfnamefont
  {Z.}~\bibnamefont {DeVito}}, \bibinfo {author} {\bibfnamefont
  {M.}~\bibnamefont {Raison}}, \bibinfo {author} {\bibfnamefont
  {A.}~\bibnamefont {Tejani}}, \bibinfo {author} {\bibfnamefont
  {S.}~\bibnamefont {Chilamkurthy}}, \bibinfo {author} {\bibfnamefont
  {B.}~\bibnamefont {Steiner}}, \bibinfo {author} {\bibfnamefont
  {L.}~\bibnamefont {Fang}}, \bibinfo {author} {\bibfnamefont {J.}~\bibnamefont
  {Bai}}, \ and\ \bibinfo {author} {\bibfnamefont {S.}~\bibnamefont
  {Chintala}},\ }in\ \href
  {http://papers.neurips.cc/paper/9015-pytorch-an-imperative-style-high-performance-deep-learning-library.pdf}
  {\emph {\bibinfo {booktitle} {Advances in Neural Information Processing
  Systems 32}}},\ \bibinfo {editor} {edited by\ \bibinfo {editor}
  {\bibfnamefont {H.}~\bibnamefont {Wallach}}, \bibinfo {editor} {\bibfnamefont
  {H.}~\bibnamefont {Larochelle}}, \bibinfo {editor} {\bibfnamefont
  {A.}~\bibnamefont {Beygelzimer}}, \bibinfo {editor} {\bibfnamefont
  {F.}~\bibnamefont {{\noopsort{buc}}{dAlch{\'e}-Buc}}}, \bibinfo {editor}
  {\bibfnamefont {E.}~\bibnamefont {Fox}}, \ and\ \bibinfo {editor}
  {\bibfnamefont {R.}~\bibnamefont {Garnett}}}\ (\bibinfo  {publisher} {{Curran
  Associates, Inc.}},\ \bibinfo {year} {2019})\ pp.\ \bibinfo {pages}
  {8024--8035}\BibitemShut {NoStop}%
\bibitem [{\citenamefont {Dagum}\ and\ \citenamefont
  {Menon}(1998)}]{dagum1998openmp}%
  \BibitemOpen
  \bibfield  {author} {\bibinfo {author} {\bibfnamefont {L.}~\bibnamefont
  {Dagum}}\ and\ \bibinfo {author} {\bibfnamefont {R.}~\bibnamefont {Menon}},\
  }\href@noop {} {\bibfield  {journal} {\bibinfo  {journal} {IEEE computational
  science and engineering}\ }\textbf {\bibinfo {volume} {5}},\ \bibinfo {pages}
  {46} (\bibinfo {year} {1998})}\BibitemShut {NoStop}%
\bibitem [{\citenamefont {Geiger}\ \emph {et~al.}(2022)\citenamefont {Geiger},
  \citenamefont {Smidt}, \citenamefont {M.}, \citenamefont {Miller},
  \citenamefont {Boomsma}, \citenamefont {Dice}, \citenamefont {Lapchevskyi},
  \citenamefont {Weiler}, \citenamefont {Tyszkiewicz}, \citenamefont {Batzner},
  \citenamefont {Madisetti}, \citenamefont {Uhrin}, \citenamefont {Frellsen},
  \citenamefont {Jung}, \citenamefont {Sanborn}, \citenamefont {Wen},
  \citenamefont {Rackers}, \citenamefont {Rød},\ and\ \citenamefont
  {Bailey}}]{e3nn_repo}%
  \BibitemOpen
  \bibfield  {author} {\bibinfo {author} {\bibfnamefont {M.}~\bibnamefont
  {Geiger}}, \bibinfo {author} {\bibfnamefont {T.}~\bibnamefont {Smidt}},
  \bibinfo {author} {\bibfnamefont {A.}~\bibnamefont {M.}}, \bibinfo {author}
  {\bibfnamefont {B.~K.}\ \bibnamefont {Miller}}, \bibinfo {author}
  {\bibfnamefont {W.}~\bibnamefont {Boomsma}}, \bibinfo {author} {\bibfnamefont
  {B.}~\bibnamefont {Dice}}, \bibinfo {author} {\bibfnamefont {K.}~\bibnamefont
  {Lapchevskyi}}, \bibinfo {author} {\bibfnamefont {M.}~\bibnamefont {Weiler}},
  \bibinfo {author} {\bibfnamefont {M.}~\bibnamefont {Tyszkiewicz}}, \bibinfo
  {author} {\bibfnamefont {S.}~\bibnamefont {Batzner}}, \bibinfo {author}
  {\bibfnamefont {D.}~\bibnamefont {Madisetti}}, \bibinfo {author}
  {\bibfnamefont {M.}~\bibnamefont {Uhrin}}, \bibinfo {author} {\bibfnamefont
  {J.}~\bibnamefont {Frellsen}}, \bibinfo {author} {\bibfnamefont
  {N.}~\bibnamefont {Jung}}, \bibinfo {author} {\bibfnamefont {S.}~\bibnamefont
  {Sanborn}}, \bibinfo {author} {\bibfnamefont {M.}~\bibnamefont {Wen}},
  \bibinfo {author} {\bibfnamefont {J.}~\bibnamefont {Rackers}}, \bibinfo
  {author} {\bibfnamefont {M.}~\bibnamefont {Rød}}, \ and\ \bibinfo {author}
  {\bibfnamefont {M.}~\bibnamefont {Bailey}},\ }\href {\doibase
  10.5281/zenodo.6459381} {\enquote {\bibinfo {title} {Euclidean neural
  networks: e3nn},}\ } (\bibinfo {year} {2022})\BibitemShut {NoStop}%
\bibitem [{\citenamefont {Harris}\ \emph {et~al.}(2020)\citenamefont {Harris},
  \citenamefont {Millman}, \citenamefont {Van Der~Walt}, \citenamefont
  {Gommers}, \citenamefont {Virtanen}, \citenamefont {Cournapeau},
  \citenamefont {Wieser}, \citenamefont {Taylor}, \citenamefont {Berg},
  \citenamefont {Smith} \emph {et~al.}}]{harris2020array}%
  \BibitemOpen
  \bibfield  {author} {\bibinfo {author} {\bibfnamefont {C.~R.}\ \bibnamefont
  {Harris}}, \bibinfo {author} {\bibfnamefont {K.~J.}\ \bibnamefont {Millman}},
  \bibinfo {author} {\bibfnamefont {S.~J.}\ \bibnamefont {Van Der~Walt}},
  \bibinfo {author} {\bibfnamefont {R.}~\bibnamefont {Gommers}}, \bibinfo
  {author} {\bibfnamefont {P.}~\bibnamefont {Virtanen}}, \bibinfo {author}
  {\bibfnamefont {D.}~\bibnamefont {Cournapeau}}, \bibinfo {author}
  {\bibfnamefont {E.}~\bibnamefont {Wieser}}, \bibinfo {author} {\bibfnamefont
  {J.}~\bibnamefont {Taylor}}, \bibinfo {author} {\bibfnamefont
  {S.}~\bibnamefont {Berg}}, \bibinfo {author} {\bibfnamefont {N.~J.}\
  \bibnamefont {Smith}},  \emph {et~al.},\ }\href@noop {} {\bibfield  {journal}
  {\bibinfo  {journal} {Nature}\ }\textbf {\bibinfo {volume} {585}},\ \bibinfo
  {pages} {357} (\bibinfo {year} {2020})}\BibitemShut {NoStop}%
\bibitem [{\citenamefont {Bosch}(2000)}]{bosch2000computation}%
  \BibitemOpen
  \bibfield  {author} {\bibinfo {author} {\bibfnamefont {W.}~\bibnamefont
  {Bosch}},\ }\href@noop {} {\bibfield  {journal} {\bibinfo  {journal} {Physics
  and Chemistry of the Earth, Part A: Solid Earth and Geodesy}\ }\textbf
  {\bibinfo {volume} {25}},\ \bibinfo {pages} {655} (\bibinfo {year}
  {2000})}\BibitemShut {NoStop}%
\end{thebibliography}
\end{document}